\documentstyle[11pt]{article}

\newcommand{\half}{\mbox{$\frac{1}{2}$}}

\newcommand{\SU}{{\rm SU}}
\newcommand{\su}{{\rm su}}

\newcommand{\eq}{\begin{equation}}
\newcommand{\eqend}{\end{equation}}
\newcommand{\eqa}{\begin{eqnarray}}
\newcommand{\nonueqa}{\begin{eqnarray*}}
\newcommand{\eqaend}{\end{eqnarray}}
\newcommand{\nonueqaend}{\end{eqnarray*}}
\newcommand{\nonu}{\nonumber \\ \nopagebreak}
\newcommand{\bma}[1]{\begin{array}{#1}}
\newcommand{\ema}{\end{array}}
\newcommand{\bc}{\begin{center}}
\newcommand{\ec}{\end{center}}

\newcommand{\Ref}[1]{(\ref{#1})}

\newcommand{\ee}[1]{\,\mbox{{\rm e}}^{#1}}
\newcommand{\ii}{i}

\newcommand{\om}{\omega}

\renewcommand{\phi}{\varphi}

\newcommand{\del}{\delta}

\newcommand{\tet}{\theta}      
      
\newcommand{\eps}{\varepsilon}

\newfam\msbfam
\batchmode\font\twelvemsb=msbm10 scaled\magstep1 \errorstopmode
\ifx\twelvemsb\nullfont\def\Bbb{\bf}
\else	\catcode`\@=11
	\font\tenmsb=msbm10 \font\sevenmsb=msbm7 \font\fivemsb=msbm5
	\textfont\msbfam=\tenmsb
	\scriptfont\msbfam=\sevenmsb \scriptscriptfont\msbfam=\fivemsb
	\def\Bbb{\relax\ifmmode\expandafter\Bbb@\else
 		\expandafter\nonmatherr@\expandafter\Bbb\fi}
	\def\Bbb@#1{{\Bbb@@{#1}}}
	\def\Bbb@@#1{\fam\msbfam\relax#1}
	\catcode`\@=\active
\fi
\newcommand{\R}{{\Bbb R}}
\newcommand{\C}{{\Bbb C}}
\newcommand{\Z}{{\Bbb Z}}

\newcommand{\f}{\frac}
\newcommand{\cA}{{\cal A}}
\newcommand{\cC}{{\cal C}}
\newcommand{\cL}{{\cal L}}
\newcommand{\cH}{{\cal H}}
\newcommand{\cF}{{\cal F}}
\newcommand{\cJ}{{\cal J}}

\newcommand{\ccr}[2]{{[} {#1},{#2} {]} }        
\newcommand{\car}[2]{{\{} {#1},{#2} {\}} }      

\newcommand{\Tra}[1]{{\rm Tr} \left({#1}\right)}          
\newcommand{\Tr}{{\rm Tr}} 
\newcommand{\TraC}[1]{{\rm Tr}_C \left({#1}\right)}        
\newcommand{\tra}{{\rm tr}_N}        
\newcommand{\trac}[1]{{\rm tr}_{N} \left({#1}\right)}         

\newcommand{\dd}{{\rm d}}

\newcommand{\hint}{\hat{\int}}

\newcommand{\hd}{\hat\dd}
\newcommand{\gl}{{\rm gl}}
\newcommand{\Dsl}{{D\!\!\!\!\slash_0}}
\newcommand{\Asl}{A\!\!\!\slash}
\newcommand{\DslA}{{D\!\!\!\!\slash_A}}
\newcommand{\hA}{\hat A}

\renewcommand{\Im}{{\rm Im}}
\renewcommand{\kill}[1]{{#1}\!\!\!\!\!\slash\;}

\begin{document}

\begin{flushright}
January 17, 1996
\end{flushright}

\vspace{.4cm}
\begin{center}
{\Large \bf Quantum Gauge Theories and Noncommutative 
Geometry}\footnote{Contribution to workshop `New Ideas in the Theory of 
Fundamental Interactions', Szczyrk, Poland 1995}\\
\vspace{0.3 cm}
{\large Edwin Langmann}\\
\vspace{0.3 cm}
{\em Theoretical Physics, Royal Institute of Technology, Stockholm, 
S--10044 Sweden}
\end{center}

\begin{abstract}

I review results from recent investigations of anomalies in fermion--Yang 
Mills systems in which basic notions from noncommutative geometry (NCG) 
where found to naturally appear.  The general theme is that derivations of 
anomalies from quantum field theory lead to objects which have a natural 
interpretation as generalization of de Rham forms to NCG, and that this 
allows a geometric interpretation of anomaly derivations which is useful 
e.g.\ for making these calculations efficient.  This paper is intended as 
selfcontained introduction to this line of ideas, including a review of 
some basic facts about anomalies.  I first explain the notions from NCG 
needed and then discuss several different anomaly calculations: Schwinger 
terms in 1+1 and 3+1 dimensional current algebras, Chern--Simons terms from 
effective fermion actions in arbitrary odd dimensions.  I also discuss the 
descent equations which summarize much of the geometric structure of 
anomalies, and I describe that these have a natural generalization to NCG 
which summarize the corresponding structures on the level of quantum field 
theory.

\end{abstract}

\section{Introduction} 

In the last 15 years or so spectacular progress in 1+1 dimensional quantum 
field theory has been made.  This was closely related to developments in 
mathematics, especially the representation theory of infinite dimensional 
Lie algebras (affine Kac--Moody algebras, Virasoro algebra).  It also was 
understood already quite some time ago that there are interesting relations 
to noncommutative geometry (NCG), a mathematical discipline due to 
Connes unifying and extending many results from geometry and functional 
analysis.  Unfortunately the case of interest to particle physics: 
3+1 dimensions, is much more complicated, and a picture similarly complete 
as in 1+1 dimensions is far from complete.  Nevertheless there has been 
progress \cite{MR,M,L1}, and there are now several examples that NCG 
provides useful mathematical tools also for higher dimensional quantum 
field theory \cite{LM1,LM2,LM3}.

In this contribution I try to explain in a simple way how NCG appears in 
quantum gauge theory and in what way it suggests efficient calculation 
tools there.  I also review \cite{L3} in which all Yang--Mills anomalies 
(Chern--Simons terms, axial anomalies, Gauss law anomalies etc.\ in all 
possible dimensions) and the descent equations connecting them 
\cite{Zumino} are generalized to NCG, and realizations of all cyclic 
cocycles \cite{C} through these generalized anomalies were found.

I should point out that there are several other recent, interesting results 
on NCG in connection with particle physics which I do not discuss here.  
Most of them concern gauge theory models on a classical (non--quantized) 
level.  My concern is how to use NCG to make progress in understanding the 
{\em quantum} structure of gauge theories with all its problems like 
divergences, anomalies etc.  Also when I refer to NCG I do not mean the 
full mathematical framework with all its deep results as e.g.\ summarized 
in a recent book by Connes \cite{C} but rather a few notions and special 
examples within that general framework which can be understood without 
going in depth into the mathematical theory.  As I see it (as far the the 
examples discussed here are concerned), NCG has provided a very useful way 
of looking at mathematical structures which appear in quantum field theory 
but were partly known to (mathematical) physicists before.  The known 
examples demonstrate that NCG provides a very powerful language for these 
kinds of problems, and new insights in quantum field theory can be hoped 
for in the future.

The plan of this paper is as follows.  In Section 2 I discuss the main 
notions from NCG needed.  I then describe the construction of fermion 
current algebras in 1+1 and 3+1 dimensions and explain how NCG naturally 
appears there (Sect.\ 3).  In Section 4 the generalization of the descent 
equations to NCG is explained, and a simple example of an anomaly 
calculation using ideas from NCG is also outlined there.  Section 5 
contains a few final remarks.

My discussion is restricted to explaining results of our group 
at KTH in Stockholm.  I would like to thank Jouko Mickelsson for a very 
pleasant collaboration.

\section{NCG: a few basic ideas and examples} 

A basis idea in NCG is to try to generalize notions and results form 
differential geometry to more general situations without underlying 
manifolds.  The strategy for this is along the following lines: considering 
some manifold $M$, it is possible to encode the geometric information about 
$M$ in terms of the algebra $\cA$ of complex valued functions on $M$ 
(product given by pointwise multiplication). Such an algebra $\cA$ is
commutative.  Reformulating notions and results in ordinary geometry 
using only properties of $\cA$ (and making no reference to the underlying 
manifold $M$), it is then often possible to generalize these through 
replacing $\cA$ by some other, in general noncommutative, algebra $\hat 
\cA$.  The algebra $\hat\cA$ is often naturally realized by operators on some 
Hilbert space $\cH$ i.e.\ is a subalgebra of the bounded operators 
$B(\cH)$ acting on $\cH$.

\vspace*{2mm}
\noindent
{\bf De Rham forms.} Let me give an example for an important concept in 
geometry generalized in NCG, namely de Rham forms.  Consider the manifold 
$\R^d$ (for simplicity) and the algebra $\cA=C_0^\infty(\R^d)$ of smooth, 
compactly supported and $\C$--valued maps $X$ on $\R^d$.  Then exterior 
differentiation $\dd$ can be defined on $\cA$ as usual, $\dd X(x) = 
\partial^i X(x)\dd x_i$ ($x=(x_1,\ldots,x_d)$, $x_i\in\R$, is a point in 
$\R^d$, and $\partial^i = \partial/\partial x_i$; summations in $i$ from 
$1$ to $d$ understood).  With that one can define de Rham forms on $\R^d$: 
Elements in $\cA$ are defined to be 0--forms, 1--forms are linear 
combinations of elements $X_0\dd X_1$, \ldots, $n$--forms linear 
combinations of elements $X_0\dd X_1\cdots \dd X_n$ ($X_i\in\cA$; the 
product here is the wedge product).  Then the (wedge) product $\om_n\om_m$ 
of a $n$ form $\om_n$ and a $m$--form $\om_m$ is naturally defined and is a 
$(n+m)$--form.  A crucial property is $\dd^2=0$ (this is how $\dd$ is 
defined), and this allows to naturally extend the definition of $\dd$ to 
all $n$--forms $\om_n$ such that $\dd\om_n$ is a $(n+1)$--form.  There is 
another useful operation on these de Rham forms, namely integration: a 
$d$--form $\om_d$ can always be written as $f(x)\dd x_1\cdots\dd x_d$ with 
$f(x)$ in $C_0^\infty(\R^d)$, thus the integral $\int\om_d$ can be 
naturally defined as $\int_{\R^d} f(x)\dd x_1\cdots\dd x_d$.  This 
definitions can be extended to all $n$--forms by setting $\int\om_n=0$ for 
all $n\neq d$.  One can then show many nice relations, like 
$\dd(\om_n\om_m) = \dd(\om_n)\om_m + (-)^n\om_n(\dd\om_m)$ (graded Leibniz 
rule), $\int\om_n\om_m=(-)^{nm}\int\om_m\om_n$ (cyclicity), $\int\dd\om_n= 0$ 
(Stokes' theorem) etc.

\vspace*{2mm}
\noindent
{\bf Generalizing de Rham forms to NCG.}
By inspection one can convince oneself that the algebra $\cA$ being 
commutative or the fact that there is a manifold $\R^d$ are not essential 
in this example.  It is therefore easy to generalize it to the following 
situation:\footnote{throughout the paper a hat indicates 
generalizations to NCG} there is some given algebra $\hat\cA$ and some 
linear mapping $\hd$ on $\hat\cA$ (mapping $\hat\cA$ to some vector space) 
such that $\hd^2=0$. Then one can construct 
the vector space $\hat\cC$ which is the linear span of monomials 
$u_0\hd u_1\cdots \hd u_n$ ($u_i\in\hat\cA$), and this is 
naturally an algebra: for example ($v_i\in\hat\cA$) 
$$ (u_0\hd u_1)(v_0\hd\hat v_1\cdots \hd v_n)
\equiv u_0\hd(u_1  v_0) \hd v_1\cdots 
\hd v_n - (u_0u_1) \hd  v_0\hd v_1\cdots 
\hd v_n 
$$ 
etc.\ (postulating a graded Leibniz rule for $\hd$ and using $\hd^2=0$).  
Moreover, setting $$\hd(u_0\hd u_1\cdots \hd u_n)\equiv 
\hd u_0\hd u_1\cdots \hd u_n$$ naturally defines $\hd$ as linear 
mapping on $\hat\cC$ such that $\hd^2=0$.  This algebra $\hat\cC$ together 
with the map $\hd$ has most of the algebraic properties of the de Rham 
forms on $\R^d$ and is what is called a {\em graded differential algebra} 
(GDA): it is a graded algebra, $\hat\cC = \bigoplus_{n=0}^\infty 
\hat\cC^{(n)}$ where $\hat\cC^{(n)}$ is the vector space generated by 
elements $u_0\hd u_1\cdots \hd u_n$ ($u_i\in\hat\cA$), $\hd$ maps 
$\hat\cC^{(n)}$ to $\hat\cC^{(n+1)}$, $\hd^2=0$, $\hat\om_n\in
\hat\cC^{(n)}$ and $\hat\om_m\in \hat\cC^{(m)}$ implies that 
$\hat\om_n\hat\om_m$ is in $\hat\cC^{(n+m)}$, the graded Leibniz rule holds 
etc.  It is thus natural to also denote elements in $\hat\cC^{(n)}$ as 
$n$--forms and $\hd$ as exterior differentiation.  In some cases is is also 
possible to naturally define an integration $\hint$ i.e.\ a linear map 
$\hat\cC\to\C$ such that cyclicity and Stokes' theorem holds.  
The triple $(\hat\cC,\hd,\hint)$ is what is called {\em cycle} by 
Connes.\footnote{see e.g.\ \cite{C} $p.183$; my definition is slightly 
more general} If $\hint$ is concentrated on some $\hat\cC^{(d)}$ 
(i.e.\ $\hint\hat\om_n=0$ for all $n$--forms except $n=d$) it is possible to 
naturally assign a dimension to a cycle {\em viz.} dimension$=d$.

\vspace*{2mm}
\noindent
{\bf Example 1.} This first example is important for Yang--Mills theory.  
Even though it starts with a noncommutative algebra, there is still an 
underlying manifold. I thus regard it as example for ordinary geometry 
which will then be generalized to NCG in Example 2 below.

Take as $\hat\cA=\cA_d$ the algebra of all complex $N\times N$ 
matrix--valued functions on $\R^d$ which are smooth and compactly supported, 
$\cA_d = C_0^d(\R^d;\gl_N)$ (the product is pointwise matrix 
multiplication, $\gl_N$ is the complex algebra of $N\times N$ matrices).  
Since matrix multiplication is not commutative, $\cA_d$ is not commutative, 
either.  However, exterior differentiation $\dd$ is naturally extended to 
$\gl_N$--valued functions such that $\dd^2=0$.  I denote as $\cC_d$ the the 
GDA constructed from these data as explained above.  There is also a natural 
integration $\int$: for all $d$--forms $\om_d = f(x)\dd x_1\cdots\dd x_d$ 
with $f(x)$ a $\gl_N$ valued function on $\R^d$, thus $\int\om_n$ can be 
defined as $\int_{\R^d}\tra(\om_n(x))$ (ordinary integration) where $\tra$ 
is the usual $N\times N$ matrix trace.  With that, $(\cC_d,\dd,\int)$ 
naturally becomes a $d$--dimensional cycle.  I will refer to it as 
(generalized) de Rham cycle.

\vspace*{2mm}
\noindent
{\bf Example 2.} Consider a separable Hilbert space $\cH$ which is 
decomposed in two orthogonal subspaces, $\cH=\cH_+\oplus \cH_-$.  The 
operator $F$ which is $\pm 1$ on $\cH_\pm$ is a grading operator, 
$F^*=F^{-1}=F$, and $\cH_\pm =
\f{1}{2}(1\pm F) \cH$ ($*$ is the Hilbert space adjoint).
Then for all $u\in B=B(\cH)$ (bounded operators on $\cH$) one can define
\eq
\label{hd}
\hd u\equiv \ii\ccr{F}{u} 
\eqend
($\ccr{a}{b}\equiv ab-ba$).  Taking $\hat\cA=B$ and this $\hd$ one can construct 
a GDA.  Thus $n$--forms $\hat\om_n$ are linear combinations of operators 
$(\ii)^n u_0\ccr{F}{u_1}\cdots \ccr{F}{u_n}$ ($u_i\in B$), and it is 
easy to see that
$
\hd\hat\om_n = \ii\left( F\hat\om_n - (-)^n\hat\om_nF \right) 
$
(check that $\hd^2=0$!). 

It is now useful to restrict this GDA as follows: Denote as $B_1$ the trace 
class operators on $\cH$ and as $B_{p}=\{a\in B| (a^*a)^{p/2}\in B_1 \}$ 
(these are the so--called Schatten classes).  I will need a few basic 
properties of these operator classes \cite{Simon}: $B_p\subset B_q$ for 
$p<q$; $a\in B$, $b\in B_p$ and $c\in B_q$ implies $ab$ and $ba$ are in 
$B_p$ and $bc$ is in $B_r$ where $1/r=1/p+1/q$; moreover, the Hilbert space 
trace $\Tra{a}$ is only defined if $a\in B_1$, and $\Tra{bc}=\Tra{cb}$ if 
$1/p+1/q \geq 1$.  We can now define the following subalgebras of $B$,
\eq
\hat\cA_d\equiv\left\{ u\in B\, \left| \, \ccr{F}{u} \in B_{d+1}\right.\right\}
 = \hat\cA_d(\cH;F)
\eqend
where $d$ is a positive integer.  
(The reason for denoting this integer as $d$ will be explained below.)

Take as $\hat\cA= \hat\cA_d$ and $\hd$ as in \Ref{hd}.  I denote as 
$(\hat\cC_d,\hd)$ the GDA constructed from these data.  An integration 
$\hint$ can then be defined as follows ($\hat\om_n\in \hat\cC_d^{(n)}$),
\eq
\label{hint}
\hat{\int} \hat\omega_n = \left\{ \bma{cc} 
\TraC{\Gamma^{d-1}\hat\omega_n} &\mbox{ for
$n= d$}
\\ 0& \mbox{ otherwise}\ema\right.
\eqend
($\Gamma^{d-1}=\Gamma$ for $d$ even and $1$ for $d$ odd) where $\Gamma$ is a 
grading operator on $\cH$ such that $F\Gamma = -\Gamma F$ and 
$\TraC{a}\equiv \f{1}{2}\Tra{a+ F aF}$ is a conditional Hilbert space 
trace\footnote{$\Tr_C$ is denoted as $\Tr'$ by Connes, see e.g.\
\cite{C} $p.293$} (check that \Ref{hint} is always 
well defined!).  Cyclicity is obvious, and also Stokes' theorem holds.  Thus 
$(\hat\cC_d,\hd,\hint)$ is a $d$--dimensional 
cycle.\footnote{$(\hat\cC_d,\hd,\hint)$ is called {\em cycle associated with 
the $(d+1)$--summable Fredholm module} by Connes, see e.g.\ \cite{C} 
$p.292$}

This latter cycle is important since it  provides a natural 
generalization of the de Rham forms discussed in Example 1: there is a 
natural embedding of $(\cC_d,\dd,\int)$ in $(\hat\cC_d,\hd,\hint)$ which I 
now describe.  As discussed in the next Section, it is this embedding which 
naturally appears in quantum field theory.

Consider the Hilbert space $\cH_d\equiv L^2(\R^d)\otimes \C^\nu 
\otimes\C^N$ where $\nu=2^{[d/2]}$ ($[d/2]$ is equal to $d/2$ 
for $d$ even and $(d-1)/2$ for $d$ odd; thus elements in $\cH_d$ are 
functions $f_{\sigma,n}(x)$ on $\R^n$ carrying a `spin index' 
$\sigma=1,\ldots,\nu$ and a `color index' $n=1,\ldots,N$).  Physically one 
can interpret this as Hilbert space of fermions on $\R^d$, and then it is 
natural to consider the free Dirac operator $\Dsl= \gamma_i 
(-\ii\partial^i)$ acting on $\cH_d$; the matrices $\gamma_i$ act on 
$\C^\nu=\C^\nu_{spin}$ and are the usual $\nu\times\nu$ $\gamma$--matrices 
obeying $\gamma_i\gamma_j +
\gamma_j\gamma_i = 2\delta_{ij}$.  {}For even $d$ there exists
an additional $\gamma$--matrix $\gamma_{d+1}$ which will be also needed.  
The Dirac operator naturally defines a grading operator 
$\eps=|\Dsl|^{-1}\Dsl$ (zero modes of $\Dsl$ are taken care of if one sets 
$|x|^{-1} x=1$ for $x\geq 0$ and $-1$ for $x<0$).  Now elements in 
$\cA_d=C_0^\infty(\R^d;\gl_N)$ naturally act on $\cH_d$ by bounded 
operators, namely by pointwise multiplication ($\gl_N$ acts on 
$\C^N=\C^N_{color}$), and 
this defines an embedding of $\cA_d$ in $B(\cH_d)$.  (For simplicity I 
will use the same symbol $X$ for an element in $\cA_d$ and the 
corresponding operator in $B(\cH_d)$.) Take $F=\eps$. 
The embedding of $\cA_d$ in $\hat\cA_d$, i.e.\
\eq
\label{embed}
X\in\cA_d \Rightarrow X\in \hat\cA_d(\cH_d;\eps)\, ,
\eqend
can be proved by an explicit (not quite trivial) calculation (see e.g.\ 
\cite{MR}).  Moreover, $\hint$ generalizes integration of de Rham forms, i.e.\
\eq
\label{ncg}
(\ii)^d \TraC{\Gamma X_0\ccr{\eps}{X_1}\cdots \ccr{\eps}{X_d}} = 
c_d \int_{\R}\trac{X_0\dd X_1 \cdots \dd X_d} 
\eqend 
$\forall X_i \in C_0^\infty (\R^d;\gl_N)$, with $\Gamma=1$ for $d$ odd and 
$\Gamma=\gamma_{d+1}$ for $d$ even.  I recently gave a proof of this quite 
nontrivial result determining also the constant 
\eq
\label{cd}
c_d= 
\f{(2\ii)^{[d/2]}2\pi^{d/2}}{d(2\pi)^d \Gamma(d/2)}
\eqend
\cite{L4}; my method of proof of this quite fundamental result in NCG 
(namely that $\Tr_C$ is a noncommutative generalization of integration of 
de Rham forms) is by direct calculation and was motivated by quantum field 
theory calculations discussed in the next Section.

I finally note that there is also a natural generalization of {\em 
partial integration} to arbitrary GDA which generalizes the notion of 
integration of de Rham forms over submanifolds to NCG \cite{L3}.  

\section{NCG and quantum field theory}  

Dimensional regularization \cite{dim} seems to be an early example for NCG 
being relevant for quantum gauge theories: it is done by formally 
considering the model on a $(4-\epsilon)$--dimensional spacetime 
$M^{4-\epsilon}$ which, of course, is not a manifold.  It is tempting to 
conjecture that a deeper understanding of $M^{4-\epsilon}$, and thus 
dimensional regularization, should be possible in the framework of NCG.  

In the following I discuss better understood examples for NCG appearing in 
quantum field theory.

\vspace*{2mm}
\noindent
{\bf Fermion quantum field theory.} It has been realized already more than 
20 years ago that several mathematical questions in quantum field theory 
(QFT) can be studied on a general, abstract level \cite{SS} which turned 
out to be similar in the spirit to NCG.  {}For many purposes, the 
appropriate QFT setting for relativistic\footnote{by this I mean that 
there is a 1--particle Hamiltonian not bounded from below} fermions can 
namely be completely characterized by, $(i)$ the Hilbert space $\cH$ of 
1--particle states, $(ii)$ a grading operator $F$ providing the splitting of 
$\cH$ in positive-- and negative energy subspaces, $\cH=\cH_+\oplus\cH_-$ 
with $F\cH_\pm =
\pm\cH_\pm$.  Then $\cH$ determines the field algebra {\em CAR}$(\cH)$ of fermion 
fields: it can be constructed as $C^*$--algebra generated by elements 
$\psi^*(f)$ and $\psi(f)=\psi^*(f)^*$, $f\in\cH$, such that 
$f\mapsto\psi^*(f)$ is linear and the canonical anticommutator relations 
(CAR) hold, $\left(\psi^*(f)+\psi(g)\right)^2=(f,g)$ (= inner product of 
$f,g\in\cH$); moreover, $||\psi^*(f)||=(f,f)^{1/2}$.  Then $F$ determines 
the appropriate representation of {\em CAR}$(\cH)$ which corresponds to 
filling the negative energy states (`Dirac sea') so as to get a positive 
many particle Hamiltonian.  This setting is appropriate for models for 
fermions in external fields characterized by a 1--particle Hamiltonian  
given by a self--adjoint operator $D$ acting on $\cH$: then the choice 
$F=|D|^{-1}D$ guarantees that the many particle Hamiltonian is positive 
definite.

In this setting one can develop a general theory of Bogoliubov 
transformations \cite{SS} (my description here is closer to Ref.\ 
\cite{R}): such a transformation $\psi^*(f)\to\psi^*(Uf)$ is always given 
by a unitary operator $U\in B(\cH)$, and there is a unitary 
many particle operator $\Gamma(U)$ implementing it, 
$\psi^*(Uf)=\Gamma(U)^*\psi^*(f)\Gamma(U)$ $\forall f\in\cH$, if and only if
\eq
\label{HS}
\ccr{F}{U}\in B_2\, ;
\eqend 
this is the well--known {\em Hilbert--Schmidt condition}.  In my notation 
above, \Ref{HS} is equivalent to $U\in\hat\cA_1$.  Thus this is an early 
example where these operator algebras $\hat\cA_d$, which now play a 
fundamental role in NCG, appear in QFT.

\vspace*{2mm}
\noindent
{\bf Current algebras in 1+1 dimensions.} Lundberg \cite{Lb} studied 
fermion currents in the general, abstract setting above (for a detailed 
presentation of this formalism I recommend \cite{CR}): Given a 
1--particle observable i.e.\ self--adjoint operator $u\in B(\cH)$, one 
tries to construct the corresponding many particle observable 
$\dd\Gamma(u)$ which should obey $\ccr{\dd\Gamma(u)}{\psi^*(f)}=
\psi^*(uf)$ $\forall f\in\cH$.  These observables $\dd\Gamma(u)$ are called 
{\em currents} since for operators $u$ given by a (possibly matrix valued) 
integral kernel, $(uf)(x)=\int \dd y\, u(x,y) f(y)$ $\forall f\in\cH$, one 
formally has $\dd\Gamma(u) = \; :\! \int\dd x\, \psi^*(x) u(x,y)\psi(y)\! 
:\;$, where the double dots symbolize normal ordering and the $\psi^*(x)$ 
are fermions fields (i.e.\ operator valued distributions such that 
$\psi^*(f)=\int\dd x\, \psi^*(x)f(x)$) as considered usually by physicists.  
Again one can prove that these currents $\dd\Gamma(u)$ exists as 
self--adjoint many particle operator if and only if the Hilbert--Schmidt 
condition $u\in\hat\cA_1$ holds.  Moreover, Lundberg found that these 
currents obey the relations
\eq
\label{current}
\ccr{\dd\Gamma(u)}{\dd\Gamma(v)}=  \dd\Gamma(\ccr{u}{v}) + \hat S_1(u,v)
\quad \forall u,v\in \hat\cA_1 
\eqend 
where\footnote{Lundberg's formula looks different but is equivalent to 
this}
\eq
\label{Lundberg}
\hat S_1(u,v)= \f{1}{2}\TraC{u\ccr{F}{v}} \, . 
\eqend
This second term here is a $\C$--number which results from normal ordering.  
In the physics literature such a term is usually referred to as {\em 
Schwinger term}.  This term \Ref{Lundberg} is a {\em 2--cocycle} from a 
mathematical point of view: it has to obey non--trivial cocycle relations 
due to basic properties of the commutator $\ccr{\cdot}{\cdot}$ i.e.\ change 
of sign under exchange of arguments (= antisymmetry) and the Jacobi 
identity.  Since normal ordering of the currents $\dd\Gamma(u)$ is not 
unique but can be modified by finite terms, $\dd\Gamma(u)\to \dd\Gamma(u) - 
b(u)$, this cocycle is only unique up to trivial terms and can be modified, 
$\hat S_1(u,v)\to \hat S_1(u,v)+ b(\ccr{u}{v})\sim \hat S_1(u,v)$ where 
$b:\hat\cA_1\to \C$ is some (smooth) linear function.  Such a trivial term 
$b(\ccr{u}{v})$ is usually denoted as {\em coboundary}.  It is thus only the 
cohomology class of this cocycle (= equivalence class under $\sim$ above) 
which is important here, at least as far as ultraviolet divergences are 
concerned; $\hat S_1$ can be shown to be nontrivial (i.e.\ there is no 
normal ordering prescription yielding no Schwinger term).  There might be 
other reasons to choose some specific normal ordering prescription, 
however.  {}For example, the normal ordering prescription leading to the form
\Ref{Lundberg} of this cocycle is quite special since only in this form its 
interpretation in the framework of NCG, as explained below, becomes 
obvious.

These general, abstract results above are directly applicable only in 1+1 
dimensions.  {}For fermions on spacetime $\R^d\times\R$ coupled to external 
Yang--Mills fields (e.g.), one is interested in the special case $\cH= 
L^2(\R^d)\otimes \C^\nu_{spin}
\otimes\C^N_{color} $ equal to our $\cH_d$ above (to be specific, I
assume that the gauge groups is $\SU(N)$ in the fundamental 
representation); if the (time independent) external Yang--Mills field is 
$A=A^i\dd x_i$, then the appropriate 1--particle operator is $D=\Dsl + \Asl$ 
where $\Asl=\gamma_i A^i$, and $F=|D|^{-1}D$.  Then an interesting class of 
observables are the generators of gauge transformations which, on the 
1--particle level, are given by elements in $u=X\in C_0^\infty(\R^d;\gl_N)$.  
Now as discussed above for $A=0$ (eq.\ \Ref{embed}) --- and one can show 
that this is true also for $A\neq 0$ --- such infinitesimal gauge 
transformation obey the Hilbert--Schmidt condition only for $d=1$, and only 
in this case Lundberg's construction applies and provides the currents 
$\rho(X)=\dd\Gamma(X)$ which, in this case, are just the time component of 
the chiral fermion current smeared by the test function $X$.  The Schwinger 
term \Ref{Lundberg} in this case becomes
\eq
\label{KM}
S_1(X,Y) = -\f{\ii}{2 \pi}\int_{\R} \trac{X\dd Y}\, .
\eqend
To show this involves a nontrivial calculation.  After our discussion in 
the last Section this is, however, trivial since $\hat S_1(X,Y)=S_1(X,Y)$ 
for $X,Y\in \cA_1$ is just is a special case of \Ref{ncg}.

Note that the relations
\eq
\label{current1}
\ccr{\rho(X)}{\rho(Y)} =\rho( \ccr{X}{Y} ) + S(X,Y) \quad \forall X,Y\in 
\cA_1
\eqend
define what is known as {\em affine Kac--Moody algebra}\footnote{to be 
precise, it is the corresponding algebra on the circle $S^1$ and not on the 
real line $\R$ as here} and \Ref{KM} is the well--known Kac--Moody cocycle.  
This algebra has been known since quite some time (see e.g.\ \cite{BH}) and 
is usually called {\em 1+1 dimensional current algebra} by physicists.  The 
abstract current algebra \Ref{current} can be regarded as a generalization 
of this to NCG.  Especially Lundberg's cocycle $\hat S_1$ corresponds is 
the noncommutative generalization of the Kac--Moody cocycle $S_1$.  To 
my knowledge, this was first pointed out in \cite{CH}.\footnote{A.L. Carey 
pointed out to me that A. Connes was aware of this before}

At this point a further remark on the role of \Ref{current1} in physical 
models might be helpful.  The fermion currents of Dirac fermions in $d+1$ 
dimensions formally are $j_\nu^a(x) = \; :\!\bar\psi(x)\gamma_\nu 
T^a\psi(x)\!:\; = \; :\! \psi^*(x)\gamma^0 \gamma_\nu T^a\psi(x)\!:\; $ 
where $x\in\R$, $\nu=0,1,\ldots, d$, $\gamma_\nu$ are the $\gamma$--matrices, 
and $T^a$ generators of the Lie algebra of the gauge group which I assume 
as $N\times N$ matrices (here I assume a Minkowski metric with signature 
$(+,-,\ldots,-)$).  {}For the present case $d=1$, these currents can be 
written in terms of the chiral fermion charges, $j_0 = \rho_+ + \rho_-$ and 
$j_1=\rho_+ - \rho_-$ where $\rho_\pm^a(x) = \; :\! \bar\psi(x)\gamma_0 
\half(1\pm \gamma_3)T^a\psi(x)\!:\; $ are the chiral currents (note that 
$\gamma_3 = \gamma_0\gamma_1$).  The 
relation \Ref{current1} now is for the chiral currents $\rho=\rho_+$, 
and one gets similar ones for $\rho_-$ but with opposite sign 
of the Schwinger term.  {}From these one obtains the commutator relations of 
the currents $j_\nu$  where a Schwinger terms appears only in the 
$j_0$-$j_1$--commutators.  (See e.g.\ \cite{LS} for a more detailed 
discussion of this.) More generally, the abstract formalism leading to
\Ref{current} provides all `hard' mathematical analysis necessary to 
construct the fermion observables interesting in several 1+1 dimensional 
quantum field theory models, at least in case of massless fermions.  In 
addition to the fermion currents discussed above, it also gives the 
energy--momentum tensor (which in special cases lead to a Virasoro algebra).  
A full construction of 1+1 dimensional QCD with massless quarks (space 
compact) in this spirit was given in \cite{LS}.\footnote{I use this 
occasion to note that some of the results in \cite{LS} (not discussed here) 
were obtained earlier by Bos using different methods \cite{B}.  I thank R.  
Jackiw for pointing this out to me.} I also note that in this case, the 
requirement of gauge invariance eliminates the above mentioned freedom in 
the normal ordering prescription for constructing the fermion currents.

\vspace*{2mm}
\noindent
{\bf Current algebras in higher dimensions.} {}From eq.\ \Ref{embed} it is 
clear that the abstract current algebra \Ref{current} is of no use in 
higher dimensions.  This is not surprising from a physics point of view: 
the Schatten ideal condition \Ref{embed} can be regarded as a precise 
characterization of ultraviolet divergences, and these are worse in higher 
dimensions.  Lundberg's construction amounts to giving a precise 
mathematical meaning to normal ordering which is the only regularization 
necessary to deal with the ultraviolet divergences occurring in 1+1 
dimensions; this is not sufficient, however, for the more severe 
divergences occurring in higher dimensions.

The analog of \Ref{current} for 3+1 dimensions was found by Mickelsson and 
Rajeev \cite{MR,M}, and their construction can be easily extended to other 
dimensions \cite{FT}.  It gives a precise mathematical meaning to 
multiplicative regularization required in higher dimensions and can be 
motivated by physical considerations: consider chiral fermions coupled to 
external Yang--Mills fields $A$ as above.  Again one takes chiral fermions 
as motivation since for them one expects a nontrivial Schwinger term (see 
below); the Schatten ideal conditions are the same for Dirac fermions, 
however, and the abstract construction will therefore apply to this case 
also.

In 1+1 dimensions the regularization of fermions currents can be chosen 
independent of $A$ and is always equivalent to the one for the simplest case 
$A=0$.  In higher dimensions this is not true and one gets currents 
$\rho(X;A)$ depending on $A$ through regularization.  Instead of the 
currents one then has to consider generators of gauge transformations 
$G(X;A)=\cL_X+\rho(X;A)$ where $\cL_X$ accounts for the action of the 
infinitesimal gauge transformation on the Yang--Mills field, 
\eq
\label{Lie0}
\cL_X 
(\cdots)(A) \equiv \left. 
\f{\dd}{\dd s}(\cdots)(A+ s( \dd X + \ii [A,X]))\right|_{s=0} 
\eqend
(Lie derivative).  
These $G(X;A)$ are usually called Gauss' law generators (their vanishing on 
physical states is Gauss' law).  

The Dirac sea corresponding to the external field $A$ is characterized by 
$F_A=|\DslA|^{-1}\DslA$ where $\DslA=\Dsl+\Asl$, and it has the important 
property that $\ccr{\eps}{F_A}\in B_{d+1}$ where $\eps=F_0$ (no external 
field) \cite{MR}.  Thus on the abstract level, it is natural to fix a 
grading operator $\eps$ and introduce the sets
\eq
Gr_d =\left\{ F=F^*=F^{-1}\in B \, \left|\, \ccr{\eps}{F}\in B_{d+1} 
\right. \right\} 
\eqend
in addition to $\hat\cA_d$. The generalization of 
\Ref{current} to higher dimensions is then 
\eq
\label{gauss}
\ccr{ \hat G(u;F) }{ \hat G(v;F) } = \hat G(\ccr{u}{v};F) + \hat S_d(u,v;F) \quad 
\forall u,v\in \hat\cA_d \, ,\quad F\in Gr_d  
\eqend 
and is an abstract generalization of the algebra of Gauss' law generators. 
{}For $d=3$ Mickelsson and Rajeev found the Schwinger term 
\eq
\label{hatc3}
\hat S_3(u,v;F) =
-\f{1}{8}\TraC{(F-\eps)\ccr{\ccr{\eps}{u}}{\ccr{\eps}{v}} } \, .
\eqend
which is a 2--cocycle due to the antisymmetry and the Jacobi identity of 
$\ccr{\cdot}{\cdot}$ as discussed above.

It is worth noting how \Ref{current} can be interpreted as a special case 
of this: for $d=1$, $\dd\Gamma(u;F)=\dd\Gamma(u;\eps)$ can be chosen 
independent of $F\in Gr_1$, thus \Ref{current} is equivalent to
\Ref{gauss} if one interprets $\hat G(u;F)=\cL_u + \dd\Gamma(u;\eps)$ where
$\cL_u (\cdots)(F) \equiv \left.  \dd(\cdots)(\ee{-\ii su}F\ee{\ii su})/
\ii \dd s \right|_{s=0}$ is the obvious generalization of the Lie 
derivative above (thus $\hat S_1(u,v;F)= \hat S_1(u,v)$ \Ref{KM} for 
$F=\eps$).

I note that the original construction of eqs.\ (\ref{gauss},\ref{hatc3})
\cite{MR,M} was in a framework quite different from the one discussed 
here.  A construction in the present framework (which perhaps is more 
closer to physicists) was given in \cite{L1}.  It was pointed out in this 
paper that the fermion currents in $\hat G(u;F)$ are not operators but 
only sesquilinear forms, and that $\ccr{\cdot}{\cdot}$ on the l.h.s.  of 
eq.\ \Ref{gauss} is not just a commutator but involves a nontrivial 
regularization depending on dimension.

As described, the motivation for this construction was to find an explicit 
field theory construction of the Gauss' law operators $G(X;A)$ for chiral 
fermions in the Hamiltonian framework.  There were cohomological arguments 
\cite{F,M2} that these should obey the relations ($d=3$)
\eq
\ccr{G(X;A)}{G(Y;A)} = G(\ccr{X}{Y};A) + S_3(X,Y;A)\quad \forall X,Y\in 
\cA_3\, , \quad A\in \mbox{{\em YM}}_3
\eqend 
({\em YM}$_3\subset \cC^{(1)}(\cA_3)$ is the set of all YM--connections) 
with a Schwinger term
\eq
\label{c3}
S_3(X,Y;A) =\f{1}{24\pi^2} \int_{\R^3} \trac{A\ccr{\dd X}{\dd Y}}  
\eqend
which can be interpreted as manifestation of the gauge anomaly in the 
Hamiltonian framework.  Thus (\ref{gauss},\ref{hatc3}) provides as a 
special case these relations if $\hat S_3(X,Y;F_A)$ is equivalent 
(cohomologous) to $S_3(X,Y;A)$.  This was shown by explicit calculation in 
\cite{LM1}.  An essential step was the insight that \Ref{hatc3} is the natural 
generalization of \Ref{c3} to NCG, and the rule for this generalization is
\eq
\label{gen}
X \to u\, , \quad
\dd X \to \ii\ccr{\eps}{u}\, , \quad
A \to F-\eps\, , \quad \int\tra \to \f{1}{c_d}\Tr_C 
\eqend
which now, after our discussion in the last Section, is quite obvious (the 
second rule is natural since, for a pure gauges $A = - \ii U^{-1} \dd U$ 
with $U\in \cA_1$ ($\SU(N)$--valued map on $\R^d$), one gets 
$\DslA=U^{-1}\Dsl U$ and thus $F_A -\eps = U^{-1}\eps U -\eps = 
U^{-1}\ccr{\eps}{U}$).  Especially for pure gauges $A$, $\hat 
S_3(X,Y;F_A)=S_3(X,Y;A)$ is just a special case of \Ref{ncg}.

\section{Noncommutative descent equations}

\vspace*{2mm}
\noindent
{\bf Yang--Mills setting and anomalies.} On a classical (non--quantized) 
level, a Yang--Mills (YM) field $A$ is given by a $\su(N)$--valued 1--form 
on a manifold $M^d$ (I assume $\su(N)$ is the Lie algebra of the gauge 
group); $M^d$ can be spacetime (Euclidean framework) or space (Hamiltonian 
framework; time fixed).  Infinitesimal gauge transformation are given by 
$\su(N)$--valued functions on $M^d$ and act on functions of $A$ by Lie 
derivative $\cL_X$ \Ref{Lie0}, especially as $\cL_X A = -\ii\dd X 
+\ccr{A}{X}$.  Then the components of the YM field strength, $\cF_{ij} = 
\partial_i A_j - \partial_j A_i +\ii \ccr{A_i}{A_j}$, can be collected in 
a 2--form, $\cF_A = -\f{1}{2}\ii \cF_{ij}\dd x^i\dd x^j$. In compact notation, 
$\cF_A=-\ii\dd(A) + A^2$, and $\cL_X \cF_A=\ccr{\cF_A}{X}$.

{}For $M^d = \R^d$, this YM setting naturally is realized in the GDA 
$(\cC_d,\dd)$ of Section 2, Example 1: $A\in\cC_d^{(1)}$ are YM fields, 
$X\in \cC_d^{(0)}$ infinitesimal gauge transformation, and then 
$\cF_A\in \cC_d^{(2)}$.  An important observation now is 
that all what was needed here are basic operations available in any GDA.  
Thus one can naturally define a generalized YM setting for arbitrary GDA 
$(\hat\cC,\hd)$: $\hat A\in \hat\cC^{(1)}$ are (generalized) YM fields, 
$u\in\hat\cC^{(0)}$ infinitesimal gauge transformations acting on functions 
of $\hat A$ by Lie derivative
\eq
\label{Lie}
\cL_u 
(\cdots)(\hat A) \equiv \left. 
\f{\dd}{\dd s}(\cdots)(\hat A+ s( \hd u + \ii [\hat A,u]))\right|_{s=0}\, , 
\eqend
and $\cF_{\hat A}=-\ii\hd(\hat A) + \hat A^2\in \hat\cC^{(2)}$ obeys then 
$\cL_u\cF_{\hat A} = \ccr{\cF_{\hat A}}{u}$ etc.  To my opinion, this is 
an insight from NCG which is fundamental in applications to quantum gauge 
theories.  It provides e.g.\ flexible regularization schemes which still 
are gauge invariant in a natural way.  I will come back to this latter and 
first continue with a discussion of anomalies in the usual YM setting (see 
\cite{Jackiw} for more details).

Anomalies are special topological terms which are local functions of such 
YM fields $A$ and infinitesimal gauge transformations $X_i$.  They are 
integrals of de Rham forms and thus do not depend on the metric of $M^d$.  
The Schwinger terms $S_1$ and $S_3$ encountered in the last section are 
examples of anomalies,\footnote{to be concrete one can assume 
$M^d=\R^d$ as before, but all formulas in the following immediately 
generalize to other manifolds $M^d$, of course}
\eqa
S_1= \int \bar\om_1^2\, ,\quad \bar\om_1^2(X_1,X_2)= 
-\f{\ii}{2\pi}\trac{X_1\dd X_2} \quad (d=1)\nonu
S_3= \int \bar\om_3^2\, ,\quad  \bar\om_3^2(X_1,X_2;A)= 
\f{1}{24\pi^2} \trac{A\ccr{\dd X_1}{\dd X_2}}\quad (d=3)  
\eqaend
(here $\int$ is integration of $\C$--valued de Rham forms).  There are many 
more examples: for all integers $n$ and $k$ such that $0\leq 2n-k-1\leq d$, 
there are special de Rham forms $\bar\om_{2n-k-1}^k \in \cC_d^{(2n-k-1)}$ 
depending on $k$ infinitesimal gauge transformations $X_i\in \cC_d^{(0)}$ 
and $A\in \cC_d^{(1)}$ giving rise to anomalies $\int \bar\om_{2n-k-1}^k$ 
obeying certain cocycle relations discussed in more detail below (the 
integration here can be over any $2n-k-1$ dimensional boundaryless 
submanifold of $\R^d$).  {}For example, $CS_{2n-1}(A)= \int \bar\om_{2n-1}^0 $ 
with (I ignore normalization constants here)
\eq
\bar\om_{2n-1}^0(A) \propto \int_0^1\dd t\, 
\trac{\left(\cF_{tA}\right)^{n-1} A}
\eqend
are the {\em Chern--Simons terms}, $Anom_{2n-2}(X,A)=  
\int\bar\om_{2n-1}^0 $ 
with
\eq
\bar\om_{2n-2}^1(X;A) \propto \int_0^1\dd t\, \sum_{n_1+n_2=n-2}
\trac{\left(\cF_{tA}\right)^{n_1}\dd X \left(\cF_{tA}\right)^{n_2} A} 
\eqend
the {\em axial anomalies}, $S_{2n-3} = \int \bar\om_{2n-3}^2 $
the {\em Gauss' law anomalies} (Schwinger terms) etc.; the lower index 
refers to the dimension where these anomalies can be relevant. 

These forms $\bar\om_{2n-k-1}^k(X_1,\ldots, X_k;A)$ are very special.  
Firstly, they are {\em $k$--chains}, i.e.\ they are linear in the the $X_i$, 
polynomial in $A$, and antisymmetric i.e.\ they change sign under 
exchanges $X_i\leftrightarrow X_j$. Moreover, there are 
interesting relations among these chains which can be conveniently written
in terms of an operator $\delta$ mapping $(k-1)$--chains to $k$--chains and
which is defined as follows, 
\eqa
\label{del}
\del f^{k-1}(X_1,\cdots,X_{k};A)=\sum_{\mu=1}^{k}(-)^{\mu-1} \cL_{X_\mu}
f^{k-1}(X_1,\cdots,\kill{X}_\mu,\cdots, X_{k};A) \nonu +
\sum_{\stackrel{\nu,\mu=1}{\nu<\mu}}^{k}(-)^{\nu+\mu}
f^{k}(\ccr{X_\nu}{X_\mu},X_1,\cdots,\kill{X}_\nu,\cdots,\kill{X}_\mu,
\cdots, X_{k};A)\, ;
\eqaend
here $\kill{X}_\mu$ means that $X_\mu$ is omitted and $\cL_X$ is the Lie 
derivative \Ref{Lie0}.  These relations start with the so--called {\em 
Chern character} {\em ch}$_{2n}(\cF_A)=\tra\left(\cF_A\right)^n$, 
which can be written as
\eq
\label{ch}
\tra\left(\cF_A\right)^n = \dd \bar\om_{2n-1}^0 
\eqend 
(exterior derivative of the Chern--Simons form).  
{}From that one derives the {\em descent equations} \cite{Zumino}
\eq
\label{desc}
\delta \bar\om_{2n-k}^{k-1} + \dd \bar\om_{2n-k-1}^k = 0 
\eqend 
for $k=0,1,\ldots ,2n$. 

These relations immediately imply $\delta\int\bar\om^{k}_{2n-k-1}=0$ which 
are the cocycle relations of the anomalies mentioned.  They are essential 
properties of anomalies \cite{Jackiw}, e.g.\ in case of the Schwinger terms 
they guarantee the validity of the Jacobi identity in the algebra of Gauss' 
law operators as discussed in the last Section.

\vspace*{2mm}
\noindent
{\bf Descent equation in NCG.} Anomalies arise as manifestations of the 
non--trivial quantum nature of quantized gauge theory models, and usually 
are understood as remnants of regularizations necessary to deal with 
divergences (for review see \cite{Jackiw}).

It is an interesting question how these terms $c_{2n-k-1}^k = 
\int\bar\om_{2n-k-1}^k$ with their rich differential geometric 
structure arise from such explicit field theory calculations.  These 
involve calculating Feynman diagrams, introducing cut--offs etc.\ and in 
the end exactly these local, metric independent anomalies appear.  How 
come? To my opinion, a hint towards a general answer to this question comes 
from the abstract current algebras described in the last Section: the 
abstract Schwinger terms $\hat S_1$ and $\hat S_3$ arise from a field 
theory construction of fermion currents but turned out to have a natural 
interpretation as generalizations of the Schwinger terms $S_1$ and $S_3$ to 
NCG and to which they becomes equal in a special case.  It thus seems that 
here, the differential geometric structure of anomalies is already present 
of the level of Hilbert space operators on which the regularizations are 
performed, and NCG seems to provide the natural language for making this 
precise.  I believe that this is true for all field theory derivations of 
anomalies.

This suggests that all anomalies $c_{2n-k-1}^k$ have similar 
generalizations to NCG. We first note that the notation of $k$--chains 
and the operator $\delta$ \Ref{del} naturally generalize to arbitrary 
GDA.  One natural conjecture for a generalization
$$
c_{2n-k-1}^k = \int \bar \om_{2n-k-1}^k(X_1,\ldots,X_k;A)  
\to \hat c_{2n-k-1}^k
$$ then is to use \Ref{gen}.  However, it is highly non--trivial to check 
that the resulting generalized anomalies indeed obey the appropriate 
cocycle relations $\delta \hat c_{2n-k-1}^k=0$ (a proof of this by direct 
calculation is already quite lengthy for $\hat S_3= \hat c_{3}^2$
\cite{L1}). 

Recently I found a generalization of the descent equations \Ref{desc} to 
every GDA $(\hat\cC,\hd)$ which then provides all generalizations of 
anomalies to NCG which automatically obey the cocycle relations. 
This result is very general since it not only applies to the GDA 
$(\hat\cC,\hd)$ of Example 2 in Section 2, but if fact to any other one 
which one might find useful in the future for performing gauge invariant 
regularizations in quantum gauge theories. 

The starting point was the observation made above that every GDA naturally 
gives rise to a generalized YM setting: $A\in \hat\cC^{(1)}$ can be 
regarded as generalized YM fields, $X_i\in \hat\cC^{(0)}$ as gauge 
transformations acting by Lie derivative \Ref{Lie} etc.  {}For general GDA, 
however, the analog of the matrix trace $\tra$ does not exist, thus I first 
tried to recast \Ref{desc} in a form not using $\tra$.  This turned out to 
be the crucial step.  Instead of the $\C$--valued de Rham forms 
$\bar\omega_{2n-k-1}^k$ I had to find descent equations for the 
$\gl_N$--valued forms $\omega_{2n-k-1}^k$ without $\tra$ taken (i.e.\ 
$\bar\omega_{2n-k-1}^k=\tra{\omega_{2n-k-1}^k}$).  These involve additional 
terms which vanish under $\tra$, and they have a natural generalization to 
all GDA.

Before writing down the generalized descent equations, I would like to 
illustrate them by giving a detailed proof for the generalization of
\Ref{ch} to NCG.  {}For $n$--forms $\omega_n$, the distinction of 
$\hd\hat\om_n$ from $\hd(\hat\om_n)$ is important in the following since I 
interpret the former as operator $\hd(\hat\om_n)+(-)^n\hat\om_n\hd$ 
acting according to the graded Leibniz rule.

{}For arbitrary GDA one can construct
{\em ch}$_{2n}(\hat A)= \left( \cF_{\hat A} \right)^n$ and write this as 
$
\int_0^1\dd t\, \f{\dd}{\dd t} \left( \cF_{t\hat A} \right)^n
$  
where $\cF_{t\hat A}= -\ii t\hd(\hat A) + t^2\hat A^2$. Then
$
\f{\dd}{\dd t} \cF_{t\hat A} = -\ii  \hd(\hat A) + 2t\hat A^2 = 
\car{-\ii\hd+t\hat A}{\hat A}
$
($\car{a}{b}\equiv ab+ba$), thus
$
\left( \cF_{\hat A} \right)^n = \int_0^1\dd t\, \sum_{m=0}^{n-1} 
\left( \cF_{t\hat A} \right)^{n-1-m}\car{-\ii\hd+t\hat A}{\hat A} 
\left( \cF_{t\hat A} \right)^{m}\, .  
$ Using now that $\cF_{t\hat A}$ commutes with $-\ii\hd+t\hat A$ (since 
$\cF_{t\hat A}=(-\ii\hd+t\hat A)^2$)  I obtain $
\left( \cF_{\hat A} \right)^n = \int_0^1\dd t\, \car{\hd+\ii t\hat A}{\nu_{2n-1}^0} 
$
where
\eq
\nu_{2n-1}^0(A) =  -\ii \sum_{m=0}^{n-1} \left( \cF_{t\hat A} \right)^{n-1-m}{\hat A} 
\left( \cF_{t\hat A} \right)^{m}\, . 
\eqend
I thus get
\eq
\left( \cF_{\hat A} \right)^n = 
\hd\left( \om^0_{2n-1} \right)  + \ii \car{\hat A}{\tilde \om^0_{2n-1}}
\eqend
where
\eq
\label{om}
\omega^0_{2n-1}= \int_0^1\dd t\, \nu^0_{2n-1}\, ,\quad \tilde \omega^0_{2n-1}= 
\int_0^1\dd t \, t \, \nu^0_{2n-1}  
\eqend
which generalizes the starting point \Ref{ch} of the descent equations 
and is valid for arbitrary GDA. (I used $\car{\hd}{\omega^0_{2n-1}} = 
\hd\left(\omega^0_{2n-1}\right)$). $\omega^0_{2n-1}$ is thus the natural 
generalization of the Chern-Simons form to NCG.  As promised, the 
additional term as compared to \Ref{ch} vanishes under $\tra$ in case of 
the GDA of Example 1 in Section 2.  More generally, it vanishes under 
partial integrations $\hint_{part}$ (all one needs for this 
is cyclicity of $\hint_{part}$).

To proceed one can calculate $\delta \omega^0_{2n-1}$.   
Similarly as above one can get by direct calculation that
$$
\delta \omega^0_{2n-1}(u;A) + \hd \left( \omega^1_{2n-2} \right) 
= \ii \ccr{\om^0_{2n-1} }{u}
-\ii \ccr{\hat A}{\tilde \omega^1_{2n-2}} 
$$
where $\omega$ and $\tilde\omega$ are given by a formula similar to \Ref{om} 
with $\nu_{2n-2}^1$ proportional to 
$$
\ii (t-1)\sum \left( \left( \cF_{t\hat A} \right)^{n_1}\hd(u)
\left( \cF_{t\hat A} \right)^{n_2}  \hat{A} \left( \cF_{t\hat A} 
\right)^{n_3}   
-  \left( \cF_{t\hat A} \right)^{n_1} \hat{A} 
\left( \cF_{t\hat A} \right)^{n_2}\hd(u)\left( \cF_{t\hat A} 
\right)^{n_3}\, \right)   ; 
$$ the sum here is over all positive integers $n_\nu$ such that 
$n_1+n_2+n_3=n-2$ (I ignore normalization constants here for simplicity; 
they can be found in \cite{L3}). This gives the generalized axial 
anomaly.  We now observe that this can be compactly written as $$
\nu_{2n-2}^1 \propto \int\dd\tet_1\dd\tet_0\left( \cF_{t\hat A} 
+ \tet_0 \hat A -\ii(t-1)\tet_1\hd(u) \right)^n 
$$
where the $\tet_\nu$ are Grassmann numbers and $\int\dd\tet_\nu$  
Grassmann integrations as usual.

At this point it is natural to conjecture that
\eq
\omega_{2n-k-1}^k \propto \int_0^1 \dd t \int\dd\tet_0 
\cdots \dd\tet_k\left( \cF_{t\hat A} 
+ \tet_0 \hat A -\ii(t-1)\sum_{\nu=1}^k \tet_\nu \hd(u_\nu) \right)^n 
\eqend 
with $\tet_\nu$ Grassmann numbers, which indeed turns out to be correct for 
$k=0,1,\ldots, n-1$. In this case one can prove  \cite{L3}
\eq
\label{d1}
\delta \omega^{k-1}_{2n-k} + \hd \left( \omega^k_{2n-k-1}\right) 
= 
\left\{ \bma{cc} \cJ\omega^{k-1}_{2n-k}  - \ii \ccr{\hat A}{\tilde 
\omega^k_{2n-k-1}} & \mbox{ for $k$ odd}\\
\cJ\omega^{k-1}_{2n-k} - \ii \car{\hat A}{\tilde 
\omega^k_{2n-k-1}} & \mbox{ for $k$ even}\ema \right.   
\eqend
(as above, the formulas for the $\tilde\omega$ are obtained from the ones 
for $\omega$ by replacing $ \int_0^1 \dd t$ through $ \int_0^1 \dd t\, t$) 
where $\cJ$ is the following operator mapping $(k-1)$--chains to 
$k$--chains,
\eqa
\label{cJ}
\cJ f^{k-1}(u_1,\cdots,u_{k};A) = \ii \sum_{\nu=1}^{k}(-)^{\nu-1}
\ccr{f^{k-1}(u_1,\cdots,{u_\nu}\!\!\!\!\!\slash\; ,\cdots, u_{k};A)}{u_\nu} \, . 
\eqaend
{}For $k=n,n+1,\ldots, 2n-1$ one gets
\eq
\omega_{2n-k-1}^{k} \propto \int_0^1 \dd t \int\dd\tet_{k}
\cdots \dd\tet_0 
\left(\, -\ii \sum_{\nu=0}^k t\tet_\nu\dd(u_\nu) + \right.
\sum_{\stackrel{\nu,\mu=1}{\nu<\mu}}^k (t^2-t)\tet_\nu 
\tet_\mu\ccr{u_\nu}{u_\mu}
\nonu \left.
+\sum_{i=\nu}^k \tet_0 \tet_\nu u_\nu \right)^n
\eqend
independent of $\hat A$, and the descent equations are somewhat simpler, 
\eq
\label{d2}
\delta \omega^{k-1}_{2n-k} + \hd \left( \omega^k_{2n-k-1}\right) 
= \cJ\omega^{k-1}_{2n-k}   
\eqend 
(for $k=2n-1$ one has $\delta \omega^{2n-1}_{0} = \cJ \omega^{2n-1}_{0} $).  
As mentioned, these equations (\ref{d1},\ref{d2}) are equal to
\Ref{desc} up to terms which vanish under partial integrations.
They imply that $\hat c^k_{2n-k-1} = \hint \omega^k_{2n-k-1}$ all are 
$k$--cocycles for any integration $\hint$ satisfying Stokes' theorem and 
cyclicity, $\delta \hat c^k_{2n-k-1}=0$. These thus are natural  
generalizations of the anomalies to NCG.

\vspace*{2mm}
\noindent
{\bf Efficient anomaly calculations.} I now want to illustrate how explicit 
field theory calculations of anomalies can be simplified using notions from 
NCG as discussed above.  In \cite{LM2} a short derivation of the axial 
anomaly in all even dimensions (Euclidean framework) was given using a 
method inspired by NCG.  Similar derivations of the axial-- and Gauss law 
commutator anomalies (Schwinger terms) using the Hamiltonian framework can 
be found in \cite{M3,LM3}.  Here I sketch a short derivation of the 
Chern--Simons term from the imaginary part of the effective action of Dirac 
fermions in all odd dimensional spacetimes (Euclidean framework).

I consider effective actions $S(A)$ of massless Dirac fermions on odd 
dimensional spacetime.  I denote as $\Dsl$ the free Dirac operator; for 
spacetime $\R^d$, e.g.\, $\Dsl=\gamma_i (-\ii\partial^i)$ as before, 
$\Asl=\gamma_i A^i$ where $A=A^i \dd x_i$ (de Rham 1--form) is the 
usual Yang--Mills field.

The effective action for fermions in the external Yang--Mills field $A$ 
formally is the determinant of the Dirac operator $\Dsl+\Asl$, or 
equivalently, the trace of its logarithm.  Due to ultraviolet and infrared 
divergences some regularization of this trace is necessary (one has to add 
a small massterm and a large momentum cutoff, e.g.).  I find it convenient 
to write this determinant as ${\rm Tr}_{reg}\int_0^1 \dd t\, \f{\dd}{\dd 
t}\, {\rm log} (\Dsl+t\Asl)$, which formally is equivalent to
\eq
\label{0}
S(A) = {\rm Tr}_{reg}\int_0^1 \dd t (\Dsl+t\Asl)^{-1}\Asl;  
\eqend
I take this as definition of the effective fermion action. 

I now {\em define} $\hA = |\Dsl|^{-1} \Asl$, $\eps=|\Dsl|^{-1}\Dsl$ and 
$\cF_{t\hA} = t\car{\eps}{\hA} + t^2\hA$ (possible zero modes of $\Dsl$ are 
taken care of by an infra red regulator not further specified here); at 
this point this can be regarded as useful notation.  Then I can write
\eq
(\Dsl+t\Asl)^{-1}\Asl = (\eps + t \hA)^{-1}\hA= (\eps + t\hA) (1+ 
\cF_{t\hA})^{-1}\hA
\eqend 
(since $(\eps + t \hA)^2 = 1+ \cF_{t\hA}$), and the imaginary part of the 
action becomes
\eq
\label{1}
\Im S(A)= \Im {\rm Tr}_{reg}\int_0^1 \dd t \sum_{m=0}^\infty (-)^m(\eps + t\hA) 
(\cF_{t\hA})^m \hA .
\eqend
This should be equal, up to a constant, to the Chern--Simons term
\eq
\label{2}
\int {\rm tr}_N \int_0^1 \dd t\, (\cF_{tA})^n A 
\eqend
where $\cF_{tA} = -\ii t\dd(A) + t^2 A^2$ (de Rham 2--form) and $d=2n+1$.  
This result is now very plausible by notation.  To prove it, however, 
requires a nontrivial calculation: namely to show that $\Im {\rm Tr}_{reg}$ 
of $\hA (\cF_{t\hA})^m \hA$ is always zero, and $\Im {\rm Tr}_{reg}$ of 
$\eps (\cF_{t\hA})^m \hA$ is always zero except for $m=n$ where it is 
proportional to $\int{\rm tr}_N (\cF_{tA})^n A$.  This latter result can be 
regarded as a mathematical result in NCG similar to \Ref{ncg}.  Note also 
that all manipulations from \Ref{0} until eq.\ \Ref{1} did not use any 
property of the operators $\Dsl$ and $\Asl$ except that they are 
self--adjoint operators on some Hilbert space, and thus everything is valid 
also for much more general situations.  The place where specific properties 
of these operators and the Hilbert space we are using enters is the step 
from
\Ref{1} to \Ref{2}. Especially only here the dimension of the underlying spacetime 
manifold enters: it is only the (regularized) trace that distinguishes the 
different dimensions, and it picks up exactly one term {\em viz.} the 
anomaly.

\section{Final comments}

In this paper I only discussed quantum field theory of fermions, thus an 
obvious question is: what about bosons? There is an analog of the abstract 
theory of fermion Bogoliubov transformations described above for bosons
\cite{R}, and in my PhD thesis I worked out a natural $\Z_2$--graded version 
of the abstract current algebra \Ref{current} containing bosons and 
fermions and extending it to include super currents mixing bosons and 
fermions \cite{GL}.  A motivation for this were supersymmetric quantum 
field theory models.  Again this formalism is sufficient only for 1+1 
dimensions, and it provides as special cases super versions of the affine 
Kac--Moody algebras and the Virasoro algebra.  In \cite{L1b} I outlined the 
construction of a boson version of \Ref{gauss} and found the boson analog 
of the Mickelsson--Rajeev cocycle \Ref{hatc3} (up to as sign it is 
identical to the fermion cocycle).  As far as I know, otherwise little is 
known about boson--analogs of results described here.

In all field theory calculations described in these paper anomalies could be 
represented as regularized trace $\Tr_{reg}$ of some Hilbert space 
operator.  A main point was that, on the level of Hilbert space operators 
(before $\Tr_{reg}$ is taken), there is an interesting algebraic structure 
which naturally can be interpreted in the framework of NCG and can be 
exploited to make calculations efficient.  Then, of course, a precise 
understanding of $\Tr_{reg}$ is necessary.  There are interesting 
mathematical properties of such regularized traces which are also related 
to NCG but were not discussed here (some discussion can be found in
\cite{LM2}).


\end{document}